\documentstyle[aps,prb,epsfig,floats]{revtex}
\begin{document}

\twocolumn[\hsize\textwidth\columnwidth\hsize\csname@twocolumnfalse\endcsname

\title{Theory of polarization enhancement in epitaxial 
BaTiO$_3$/SrTiO$_3$ superlattices}
\author{J. B. Neaton and K. M. Rabe}
\address{Department of Physics and Astronomy, Rutgers University,\\
Piscataway, New Jersey 08855-0849}
\date{\today}
\draft
\maketitle
\begin{abstract}
The spontaneous polarization of epitaxial 
BaTiO$_3$/SrTiO$_3$ superlattices is studied as a function of composition 
using first-principles density functional theory 
within the local density approximation.  
With the in-plane lattice parameter fixed to that of 
bulk SrTiO$_3$, the computed superlattice polarization
is enhanced above that of bulk BaTiO$_3$ for 
superlattices with BaTiO$_3$ fraction larger than 40\%. 
In contrast to their bulk paraelectric character, the SrTiO$_3$ 
layers are found to be {\it tetragonal and polar}, possessing 
nearly the same polarization as the BaTiO$_3$ layers. General electrostatic 
arguments elucidate the origin of the polarization in the SrTiO$_3$ layers,
with important implications for other ferroelectric nanostructures.
\end{abstract}

\pacs{PACS: 64.70.Nd, 68.65.Cd, 77.22.Ej, 77.84.Dy}

\narrowtext

]

Experimental capabilities now allow layer-by-layer epitaxial growth
of perovskite-based oxides, facilitating the exploration of a wide range
of artificial materials inaccessible by conventional 
solid-state synthesis. Recently, short-period BaTiO$_3$/SrTiO$_3$ superlattices, 
with modulation lengths down to three perovskite layers, 
have been grown on SrTiO$_3$ substrates using pulsed-laser deposition 
(PLD)\cite{shimuta} and reactive molecular beam epitaxy (MBE).\cite{schlom2}
These superlattices are reportedly free of 
dislocations and coherently matched to the substrate, 
implying misfit strains of over 2\% in the BaTiO$_3$ layers.
Strains of this magnitude are expected to increase
the polarization,\cite{pertsev1,us} and the superlattice geometry
preserves the high-strain state and prevents
relaxation of the BaTiO$_3$ (BT) layers. But since the BT 
layers are interleaved with SrTiO$_3$ (ST), which has zero spontaneous 
polarization in the bulk, the degree to which the overall 
polarization might be enhanced in the superlattice remains in question.

In this Letter, we address this issue directly through study
of a series of ideal short-period BT/ST superlattices with
varying composition using first-principles density functional calculations.
Our calculations show that these superlattices are ferroelectric, 
and predict that some possess polarizations significantly larger 
than bulk BT. An atomic-level analysis also reveals a significant 
{\it non-zero polarization} and strain in the ST layers, a direct 
result of internal polarizing fields originating in the BT layers. 
These results, consistent with recent measurements,\cite{shimuta,schlom2} 
illustrate the importance of electrical boundary conditions 
for sustaining polarization in these and other nanoscale ferroelectric materials.

To predict the ground state structure and polarization in BT/ST superlattices
with various numbers of BT layers, we use density functional theory 
within the local density approximation (LDA).\cite{hks} A plane-wave 
basis set and projector-augmented wave potentials\cite{paw}
as implemented in the Vienna {\it ab initio} 
Simulations Package (VASP)\cite{kresse1,kresse2} are employed.
We consider period-five superlattices epitaxially grown on ST,
and construct 1$\times$1$\times$5 supercells, designated 4/1, 3/2, 2/3, and 1/4,
where the notation $x$/$y$ refers to $x$ perovskite 
layers of BT and $y$ layers of ST.
The underlying ST substrate is treated implicitly by 
constraining the in-plane lattice constant of each superlattice to 3.863~{\AA},
the value we calculate for the equilibrium lattice constant of cubic ST.
The computed in-plane lattice constant for bulk tetragonal BT
(3.945~{\AA}) is likewise slightly smaller than experiment, a well-known 
artifact of the LDA. The resulting mismatch between the theoretical lattice constants
is 2.1\%, in excellent agreement with that observed experimentally.
The ions within each supercell are allowed to relax toward equilibrium along [001],
within space group $P4mm$ (point group C$_{4v}$),
until the Hellmann-Feynman forces are less than 10$^{-3}$ eV/{\AA}; 
the total energy with respect to the normal (or $c$-axis) lattice parameter 
of each superlattice is minimized concurrently. 
Brillouin zone (BZ) integrations are performed with a 6$\times$6$\times$2 Monkhorst-Pack mesh.
A 44 Ry plane-wave cutoff is used for all calculations.
For Berry-phase polarizations,\cite{ksv} we find that 6 $k$-points/string along [001]
and 6 strings in the irreducible wedge provide well-converged results.

All calculations are performed under periodic, ``short-circuit'' boundary 
conditions, equivalent to a metallic substrate with the lattice 
constant of ST and full charge compensation on top and bottom ``electrode'' 
layers. Canting of polarization toward [111] is found to be energetically 
unfavorable, although we defer to a future study a full investigation of the possibility of 
zone-boundary octahedral rotations, such as those present in ST at low 
temperature.\cite{fleury,sai}


\begin{table}
\caption{Structural parameters computed for superlattices 
with in-plane lattice constant $a=3.863$~{\AA},
and with $l_{\rm ST}$ layers of SrTiO$_3$ and $l_{\rm BT}$ 
layers of BaTiO$_3$ ($\alpha$ is their ratio). 
$\langle c/a \rangle_{\rm BT}$,$\langle c/a \rangle_{\rm ST}$,
and $\langle c/a \rangle_{\rm I}$ are the average local c/a
parameters within BT, ST, and interface layers for each superlattice,
as described in the text.  P$_0$=24.97~$\mu$C/cm$^2$ is the computed value for bulk 
tetragonal BaTiO$_3$. Blank entries indicate layers  
absent for a given $\alpha$.}
\label{tab:struct}
\vskip 0.1cm
\begin{tabular}{ccccccc}
$l_{\rm BT}/l_{\rm ST}$ & $\alpha$ & $c/a$ & $\langle c/a \rangle_{\rm BT}$ &
$\langle c/a \rangle_{\rm ST}$ &
$\langle c/a \rangle_{\rm I}$ & P/P$_0$\\
\hline
5/0 & 0        & 5.3426 & 1.0685 &        &        & 1.570\\
4/1 & 0.25     & 5.2502 & 1.0609 &        & 1.0338 & 1.435\\
3/2 & 0.667    & 5.1820 & 1.0566 & 1.0077 & 1.0306 & 1.249\\
2/3 & 1.5      & 5.1181 & 1.0526 & 1.0055 & 1.0272 & 1.000\\
1/4 & 4        & 5.0510 &        & 1.0018 & 1.0227 & 0.520\\
0/5 & $\infty$ & 5.0000 &        & 1.0000 &        & 0.000\\
\hline
\end{tabular}
\end{table}

The $c/a$ lattice parameters computed for 
the four period-5 superlattices, as well as those for bulk ST (0/5) and
bulk strained BT (5/0), are provided in Table I.
With the in-plane lattice parameter fixed to the computed ST lattice
constant, the $c/a$ ratio of BT expands to 1.0685 
($\times$5 layers = 5.3426 for the 5/0 superlattice), in agreement
with the experimentally determined Poisson ratio,\cite{shimuta,schlom2}
and the $c/a$ ratio and cell volume decrease by about 2\% per additional SrTiO$_3$ layer. 
The decrease in $c/a$ results in a concomitant reduction in
the polar distortion within each layer. This is quantified by
dividing each superlattice into five Ti-centered unit cells (or layers), 
with A cations at the corners, and examining local displacements from
the pseudocubic positions within each layer. The individual perovskite layers are
labeled as BT, if bounded by two BaO layers; ST, if bounded by SrO layers; 
or I, if bounded by one BaO and one SrO layer (one of the two
interface layers within the supercell). The distance between neighboring A cations
can be regarded as a local $c/a$; the average values of
this ratio for each layer type, $\langle c/a \rangle$, appear Table I.
The large $\langle c/a \rangle_{\rm BT}$ are expected, based on
the considerable in-plane strain.\cite{us} More surprisingly, 
the average local $c/a$ of the ST layers deviates slightly from unity: 
the ST unit cell is evidently {\it expanded} within the superlattice
relative to the bulk. The strain is
accompanied by relative displacements within the ST layers, where in particular
we find the equatorial oxygens (Wyckoff positions 2c) to move significantly
offsite. 

The observed structural trends are reflected in the computed superlattice 
polarizations, also given in Table I. Despite the presence of 
one SrO layer, the polarization of the 4/1 superlattice is 
predicted to be considerably enhanced 
over that of BT under standard conditions in the bulk, and also
larger than expected if the ST were nonpolar. Replacing one BaO layer 
with SrO reduces the polarization by only about
10\%, while a simple scaling of the polarization by
the volume fraction of BT would result in a 20\% reduction.
The 3/2 and 2/3 superlattices are found to retain 80 and 64\% of the 
polarization of pure strained BT, again surpassing the scaled
values of 60\% and 40\%. The simple scaling argument is inadequate
because it assumes bulk behavior for the ST and strained BT layers, 
and neglects any electrostatic coupling between them. Internal polarizing fields
play an essential role in determining the properties of these superlattices, as has been
noted previously for other superlattice systems.\cite{shen,sepliarsky}

\begin{figure}
\begin{center}
   \epsfig{file=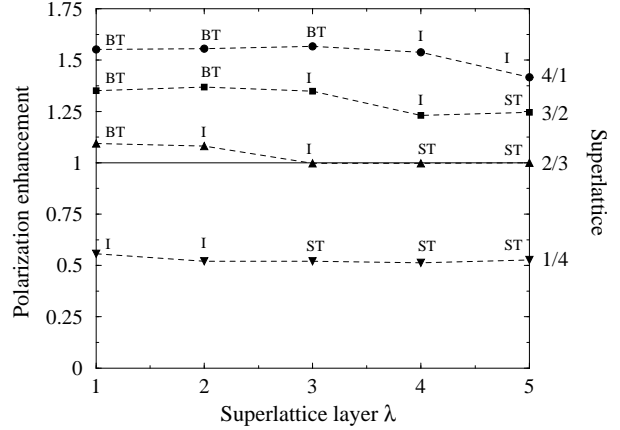,width=3.2in}
\end{center}
\caption{Local polarization enhancement (P$_\lambda$/P$_0$) 
by layer $\lambda$ for each superlattice. Each layer
is labeled BT, ST, or I as discussed in the text. As before,
P$_0$=24.97~$\mu$C/cm$^2$.\protect}
\label{fig:fig1} 
\end{figure}

To gain insight into the electrostatic coupling between layers, 
we decompose the total superlattice polarization into contributions from 
the individual Ti-centered perovskite layers defined above.
(An alternate but equally valid A cation-centered local cell choice
does not result in significant changes in our analysis.)
The local polarization $P_\lambda$ of cell $\lambda$ 
is expressed as a function of displacements within that cell as
\begin{equation}
P_\lambda \cong \sum_i \frac{\partial P}{\partial u_{\lambda 0i}} 
(u_{\lambda i} - u_{\lambda 0i})={1\over\Omega}\sum_i Z^*_i \Delta u_{\lambda i},
\end{equation}
where $\Delta u_{\lambda i}$ is the displacement of ion $i$ in unit
cell $\lambda$, $Z^*_i$ is the dynamical effective charge of ion $i$,
and $\Omega$ is the volume of the superlattice supercell.
(All polarizations and displacements are along [001].)
Since there are two different apical O ions (corresponding
to Wyckoff positions 1b) associated with the buckled AO plans 
bounding the cell, we average their contribution to each 
cell, i.e., $\Delta \bar{u}_{\lambda,{\rm 1b}}$=$1\over 
2$($\Delta u_{\lambda+1,{\rm 1b}}+\Delta u_{\lambda,{\rm 1b}})$.
A similar prescription has previously been used in a study of domain
walls in PbTiO$_3$.\cite{bernd} The dynamical effective charges 
are evaluated by finite differences, in a reference structure with atomic 
displacements identically zero within every layer;
their values will appear elsewhere.\cite{us2}

In Figure 1 the local polarization is shown as a function of layer for 
each superlattice. The most conspicuous feature of these 
profiles that the local polarization is nearly {\it constant} 
throughout each superlattice, minimizing electrostatic energy costs 
associated with the build-up of polarization charge $\nabla\cdot${\bf P} 
at the interfaces. Not only are the ST layers
polarized, but their local polarizations are close to those of the BT layers. 
Notably, the ST polarization, induced by the presence of the BT layers, is significant 
even in the 1/4 superlattice with just a single BaO layer, and even in
the absence of in-plane strain. The polarization in the BT layers, 
on the other hand, is reduced and close in magnitude to that in 
the ST layers. If the local polarizations of the BT and ST layers were
taken as those of the bulk, then the electric fields within the constituent 
BT and ST layers would be finite and have opposite signs. The field within the BT layers 
would oppose the polarization, and that in the ST layers
would polarize the nominally nonpolar layers. 

To examine this in more detail, we model the system by two slabs 
of linear dielectric media, having dielectric constants 
$\epsilon_{\rm b}$ and $\epsilon_{\rm s}$, in a parallel plate geometry.
Short-circuit boundary conditions require the electric fields within each slab 
to be related by 
${\mathcal{E}}_{\rm b} l_{\rm BT}$=$-{\mathcal{E}}_{\rm s} l_{\rm ST}$, 
where $l_{\rm BT}$ and $l_{\rm ST}$ correspond to the number of 
layers of BT and ST, as in Table I. Setting their electric displacements
$D = {\mathcal{E}} + 4\pi P$ to be equal, and 
using $P_{\rm s}=\chi_{\rm s}{\mathcal{E}}_{\rm s}$
and $P_{\rm b}=\chi_{\rm b}{\mathcal{E}}_{\rm b}+P_{5/0}$, the
polarization of each of the two slabs can be expressed in terms of the
dielectric constants of the two materials, the initial spontaneous polarization, $P_{5/0}$,
of strained BT, and their respective thicknesses.  
The average of the total slab polarization $\bar{P}$, weighted by the number of layers
of each constituent, then becomes
\begin{equation}
\bar{P} ={P_{5/0}\over 1 + \alpha ({\epsilon_{\rm b}/\epsilon_{\rm s}})},
\end{equation}
where $\alpha={l_{\rm ST}/l_{\rm BT}}$ and $P_{\rm 5/0}$ is the computed
polarization of strained BT, here 39.20~$\mu$C/cm$^2$ (as in Table I).
This macroscopic formula, valid to linear order in the dielectric constants, 
can be seen to explain the composition dependence of the superlattice 
polarization. In Fig.~2 we plot the total polarization of each superlattice, calculated
from first principles, against $\alpha$. We then perform a one-parameter fit to this
data, obtaining the dashed curve in Fig.~2. Notice that the curve 
approximates the data remarkably well, even though our expression for the polarization
is derived from macroscopic electrostatics and the superlattice layers
are atomically thin. Thus to maximize the superlattice polarization,
one must increase the fraction of BT layers (while preserving the high
strain state), consistent with the recent 
measurements of Shimuta {\it et al.}\cite{shimuta} 
It is particularly striking that even for superlattices containing 
as little as 40\% BT, one can expect to achieve a polarization as 
large as that of bulk BT.

\begin{figure}
\begin{center}
   \epsfig{file=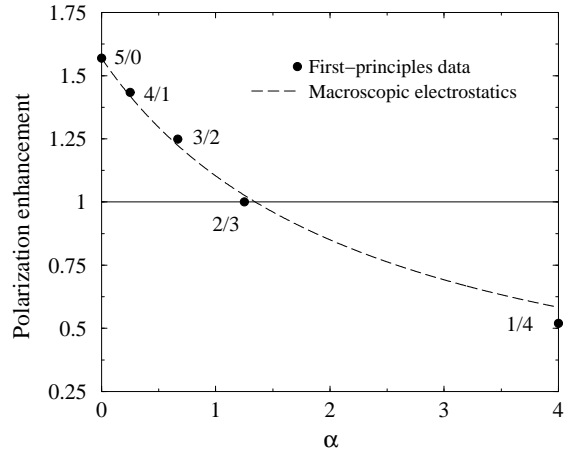,width=2.9in}
\end{center}
\caption{Polarization enhancement ($\bar{P}/P_0$), computed
from first principles, as a function of $\alpha={l_{\rm s}/l_{\rm b}}$ 
for each superlattice (filled circles). 
${\epsilon_{\rm b}/\epsilon_{\rm s}}$=0.4229 provides the best fit
of our calculations to Eq.~(2).\protect}
\label{fig:fig2}
\end{figure}

The novel properties of short-period superlattice materials are, quite generally, 
expected to arise from three separate effects. The first is the sustainability 
of large strains in sufficiently thin lattice-mismatched layers. This can
produce a significant change in properties from
the bulk at standard conditions, manifest here in the enhanced
polarization of the BT layers. We also expect electrostatic effects 
if the two materials have different susceptibilities or a
{\it polarization mismatch} (i.e., different bulk polarizations),
as in the present case. Indeed, internal fields can largely determine polarization and 
structural properties, resulting in a 
uniform polarization throughout this chemically inhomogeneous BT/ST system.
Finally, we anticipate changes in properties associated with a 
high concentration of interfaces, where the bonding and structure will
in general depart from that of the interior of the layers or the
bulk. While in this system the interfaces are relatively gentle,
apparently playing a minor role in the structure and polarization,
they may be important or even dominant in other superlattice
structures, especially if the thickness of the individual constituent 
layers is only a few unit cells.

In conclusion, we have shown, using a series of first-principles calculations, 
that significant polarization enhancement can be achieved in perovskite 
oxide superlattices. This enhancement arises from the combined effects of strain, 
induced in the BT layers by the epitaxial growth, and internal electric
fields, associated with the superlattice geometry, which
polarize the ST layers. The induced polarization observed here
complements a previous prediction of ferroelectricity in ST layers 
under epitaxial stress.\cite{pertsev1,pertsev2} 
Our analysis reveals important physical factors that influence
the behavior of the superlattice system for given constituent layers
and thickness, which will aid in the prediction of properties of a
wider class of systems and provide a valuable guide for the design of
artificially-structured materials.

\acknowledgments

We thank C. H. Ahn, M. H. Cohen, X. Q. Pan, D.~G.~Schlom,
and D. Vanderbilt for valuable discussions.
This work was supported by NSF-NIRT DMR-0103354 and ONR N00014-00-1-0261.


\end{document}